\documentstyle[epsfig]{elsart}

\def\slashchar#1{\setbox0=\hbox{$#1$}           
   \dimen0=\wd0                                 
   \setbox1=\hbox{/} \dimen1=\wd1               
   \ifdim\dimen0>\dimen1                        
      \rlap{\hbox to \dimen0{\hfil/\hfil}}      
      #1                                        
   \else                                        
      \rlap{\hbox to \dimen1{\hfil$#1$\hfil}}   
      /                                         
   \fi}                                         %

\newcommand{\ifig}[1]{\mbox{\epsfig{file=#1,height=10cm,width=14cm}}}

\def\bc{\begin{center}}
\def\ec{\end{center}}
\def\be{\begin{equation}}
\def\ee{\end{equation}}
\newcommand{\chilim}{\lim_{m \rightarrow 0}}
\newcommand{\ba}{\begin{eqnarray}}  
\newcommand{\ea}{\end{eqnarray}}

\newcommand{\nn}{\nonumber}

\newcommand{\RI}{\mbox{\scriptsize RI}}

\newcommand{\HV}{\mbox{\scriptsize HV}}
\newcommand{\NDR}{\mbox{\scriptsize NDR}}

\newcommand{\msbar}{\overline{\mbox{MS}}}

\newcommand{\bea}{\begin{eqnarray}}
\newcommand{\eea}{\end{eqnarray}}

\newcommand{\as}{\alpha_s} 
\newcommand{\J}{\hat{J}}
\newcommand{\W}{\hat{   W}}
\newcommand{\U}{\hat{ U}}
\newcommand{\gammaz}{\hat{\gamma}^{(0)T}}

\begin{document}
\vspace{-1.0cm}
\rightline{BUHEP-99-23}
\rightline{FTUV/99-62}
\rightline{IFIC/99-65}
\rightline{ROME1-1265/99 }
\begin{frontmatter}
\title{Renormalization Group Invariant  Matrix Elements of
        $\Delta S = 2$ and $\Delta I = 3/2$ 
       Four-Fermion Operators without Quark Masses}
\vspace{-1.0cm}
\author{A.~Donini}
\address{Dep. de Fisica Teorica, Univ. Autonoma de Madrid,\\ 
         Fac. de Ciencias, C-XI, Cantoblanco, E-28049 Madrid, Spain.}
\author{V.~Gim\'enez}
\address{Dep. de Fisica Teorica and IFIC, Univ. de Valencia,\\ 
         Dr. Moliner 50, E-46100, Burjassot, Valencia, Spain.}
\author{L.~Giusti}
\address{Department of Physics, Boston University\\
         Boston, MA 02215 USA.}
\author{G.~Martinelli}
\address{Dip. di Fisica, Univ. di Roma ``La Sapienza'' and\\
INFN, Sezione di Roma, P.le A. Moro 2, I-00185 Roma, Italy.}
\begin{abstract} 
We introduce  a new parameterization of four-fermion  operator matrix
elements which
does not involve quark masses and thus allows a reduction of   
systematic uncertainties. In order to simplify the matching
between  lattice and continuum renormalization schemes, we
express our results in terms of   renormalization group invariant
$B$-parameters which  are renormalization-scheme and scale 
independent. As an application of our proposal, matrix elements of  $\Delta I=3/2$ 
and SUSY $\Delta S =2$  operators  have been computed.
The calculations have been performed using the tree-level improved
Clover lattice action at two different values of the strong coupling constant
($\beta=6/g^2=6.0$ and $6.2$), in the quenched approximation. 
Renormalization constants and mixing coefficients of lattice operators 
have been obtained non-perturbatively. Using lowest order $\chi$PT,
we also obtain 
$\langle \pi \pi| O_7|K \rangle^{\NDR}_{I=2} 
= (0.11\pm 0.02)$~GeV$^4$ and 
$\langle \pi \pi| O_8|K \rangle^{\NDR}_{I=2} 
= (0.51\pm 0.05)$~GeV$^4$ at $\mu=2$~GeV.
\end{abstract}
\end{frontmatter}
\centerline{PACS: 11.15.H, 12.38.G, 14.40.Aq and 12.15.Hh}
\newpage
\clearpage 
\section{Introduction}
\label{sec:intro}
\vspace{-0.5cm}
Since  the original proposals of using lattice QCD to study hadronic 
weak decays~\cite{CMP}--\cite{bernardargo}, 
substantial theoretical and numerical progress has been made:
the main theoretical aspects of the renormalization of composite four-fermion
operators  are fully understood~\cite{boc,shape1}; the calculation of  $K^0$--$\bar K^0$ 
mixing,  expressed in terms of the so-called renormalization group
invariant $B$-parameter $\hat B_K$, has reached a level of accuracy which is unpaired by any
other approach~\cite{ksgp}--\cite{Sharpe96}; 
increasing precision has also been gained in the determination of the
electro-weak penguin amplitudes necessary to  the prediction
of the CP-violation parameter 
$\epsilon^\prime/\epsilon$~\cite{bs}--\cite{NoiDELTAS=2}; 
attempts to compute the matrix element of 
the QCD penguin operator $O_6$ exist~\cite{soni}. Finally, matrix  elements of  $\Delta S=2$ operators which are relevant 
to study FCNC effects in SUSY models have been also  computed~\cite{NoiDELTAS=2,ds2susy}.
\par Following the common lore,  matrix elements of  weak four-fermion operators are 
given in terms  of the  so-called $B$-parameters which measure 
the deviation  of their values from those obtained in the Vacuum 
Saturation Approximation (VSA).  A classical example is provided by the 
matrix element of the $\Delta S=2$ left-left operator  $O^{\Delta S=2}=
 \bar s \gamma_\mu (1- \gamma_{5} ) d  \; \bar s \gamma^\mu (1- \gamma_{5} ) 
 d$  relevant to the prediction of   the CP-violation parameter $\epsilon$
\be  \langle  \bar K^{0} \vert   O^{\Delta S=2}  \vert K^{0} 
\rangle = \frac{8}{3} M_{K}^{2} f_{K}^{2} B_{K} \ . \ee
VSA values and $B$-parameters are also used for matrix elements 
of
$\Delta S=1$ operators entering  
$\epsilon^\prime/\epsilon$, in 
particular $O_{6}=\bar s_\alpha\gamma_\mu(1-\gamma_5)d_\beta\sum_q 
\bar q_\beta 
\gamma_\mu (1+\gamma_5) q_\alpha$ and $O_{8}= 3/2 
\bar s_\alpha\gamma_\mu(1-\gamma_5)d_\beta\sum_q e_q \bar q_\beta 
\gamma_\mu (1+\gamma_5) q_\alpha$~\cite{epseM,epseR}
\bea\label{eq:Bbrutti}
\langle \pi\pi|  O_6(\mu) | K\rangle_{I=0} & = & - 4 
\left[\frac{M^2_{K}}{m_s(\mu)+ m_d(\mu)}\right]^2 (f_K - f_\pi) 
\; B_6(\mu)\; ,\nonumber\\
\langle \pi\pi| O_8(\mu) |K\rangle_{I=2} & = &
\sqrt{2} f_\pi \left[
\left(\frac{ M^2_{K} }{ m_{s}(\mu) + m_d(\mu) }\right)^{2}\right.\nonumber\\ 
& -& \left.\frac{1}{6}\left(M_K^2 - M_\pi^2\right)\right] \;
B^{(3/2)}_8(\mu)\; .
\eea
Contrary to  $\langle  \bar K^{0} \vert   O^{\Delta S=2} \vert K^{0} 
\rangle$  and $\langle \pi\pi|  O_6 | K\rangle_{I=0}$
which vanish in the chiral limit,
the matrix element of the left-right operator   
$\langle \pi\pi| O_8 |K\rangle_{I=2}$ 
remains  finite. For this reason, the dependence of this amplitude
on quark masses is expected to be smooth because it only 
enters  in  higher orders of the chiral expansion 
\be \left( \frac{ M^2_{K} }{ m_{s} + m_d }\right)^{2} \sim
C_{1} + C_{2}\;(m_{s} +m_{d})  +{\cal O}(m_{s}^{2}) \, , \label{eq:ecl} \ee
where $C_{1,2}$ are constants, independent of quark masses. 
\par Quark masses, however,  appear explicitly in   
(\ref{eq:Bbrutti}).  Since in the VSA (and in the $1/N$ 
expansion~\cite{oon}) the expression of the matrix elements 
is quadratic  in $m_{s} + m_d$, predictions for the physical 
amplitudes  are  heavily affected by the specific value 
taken for 
this quantity. Contrary to $f_{K}$, $M_{K}$, 
etc.,    quark masses  are not  directly measured by experiments and the present accuracy  
in their determination is still rather  poor~\cite{qmasses}.
Therefore,  the ``conventional'' parameterization~(\ref{eq:Bbrutti})  
introduces  a large systematic  uncertainty in the prediction of the 
physical amplitudes 
of $\langle  O_6\rangle_{I=0} $ and 
$\langle  O_8 \rangle_{I=2} $ (and of any other left-right 
operator).  Moreover, whereas for $O^{\Delta S=2}$ we introduce $\hat B_{K}$ 
as an alias of the matrix element,  by using~(\ref{eq:Bbrutti}) we 
replace each of the matrix elements with 2 unknown quantities, i.e. 
the $B$-parameter and $m_{s} + m_d$.   Finally, in many 
phenomenological analyses, 
the values of the $B$-parameters of $\langle  O_6\rangle_{I=0}$ and 
$\langle  O_8 \rangle_{I=2}$
and of the quark masses are taken by independent lattice 
calculations, thus increasing the spread of the theoretical 
predictions~\footnote{ We will discuss the correlation between the 
value of the $B$-parameter and the quark masses in 
sec.~\ref{sec:numeri}.}.
All this can be avoided in the lattice approach, where matrix elements can be computed 
from first principles. 
\par In this paper, we  propose a new   parameterization of   matrix elements in terms of well known
 experimental  quantities, without any reference  to the VSA and therefore 
 to  the strange (down) quark mass. 
This results in  a determination of  physical amplitudes with  smaller systematic errors.
As an application of our proposal we have reanalyzed the lattice correlation functions 
considered in~\cite{NoiDELTAS=2} to estimate   matrix elements of 
$\Delta I=3/2$ operators and of the operators of the  most 
general $\Delta S =2$ Hamiltonian. By comparing the results of the 
present study with those of ref.~\cite{NoiDELTAS=2} we show all the 
advantages of the new parameterization. We  give the results for  operators renormalized non-perturbatively
in the RI (MOM) scheme~\cite{DELTAS=2,NoiDELTAS=2,NP,nonper}. 
\par We also introduce a Renormalization Group 
Invariant (RGI) definition of  matrix elements (and Wilson  coefficients).  
This definition  generalizes to  an arbitrary basis a 
concept which has been  adopted very successfully for   the  $K^{0}$--$\bar K^{0}$ mixing 
amplitude,  which is usually written in terms of the RGI $B$-parameter, $\hat B_{K}$.
 With  our definition,  Wilson coefficients and operators
are renormalization-scale and scheme independent.   This will 
hopefully avoid some  confusion existing in the literature.  This confusion is generated by
the fact that (perturbative)  Wilson  coefficients and  
(non-perturbative) matrix elements have been  computed using different 
techniques, regularizations, renormalization schemes (different versions 
of NDR and 
HV, DRED,  RI-MOM with different external states) and renormalization scales. 
In this way, the effective Hamiltonian is splitted in terms  which are individually scheme and 
scale independent.
\par
The remainder of the paper is organized as follows: in sec.~\ref{sec:definitions} we define  
operators  and  matrix elements considered in the present study and 
introduce the new parameterization of the matrix elements;  in sec.~\ref{sec:RGI} we give the
RGI definition for a generic operator basis;  in sec.~\ref{sec:numeri} we 
present our new numerical results and compare them with those 
obtained with the ``conventional'' definition of the $B$-parameters;
in sec.~\ref{sec:fine} we give our best estimates of the matrix elements and 
in  sec.~\ref{sec:conclusions} we present our conclusions.
All details concerning  the non-perturbative renormalization of  
lattice operators and  the extraction of matrix elements from 
correlation functions are not discussed here since they can 
be found in refs.~\cite{DELTAS=2,NoiDELTAS=2}. In particular, in 
ref.~\cite{NoiDELTAS=2}, 
the same set of lattice data was analyzed. We compare the results of 
the present study to those of this reference.    
\vspace{-0.5cm}
\section{Matrix elements without quark masses}
\label{sec:definitions}
\vspace{-0.5cm}
In this section we introduce the  notation   and define operators and  matrix elements 
used in this paper. The new parameterization, that does not involve any reference to quark masses, 
is defined here.   An alternative parameterization which 
(for reasons explained below)  has not been used in our numerical 
analysis,  but may be useful in the future, is also considered in this section.   
\begin{itemize}
\item \underline{ $\Delta S =2$ operators:}\\
The analysis of $K^0-\bar K ^0$ mixing with the most
general $\Delta S =2$ effective Hamiltonian requires   
the knowledge of the matrix elements
$\langle\bar K^0| O_i |K^0 \rangle$ of the following operators~\cite{susy}--\cite{strumioski} 
\bea 
O_1 &=& \bar s^\alpha \gamma_\mu (1- \gamma_{5} ) d^\alpha \ 
\bar s^\beta \gamma_\mu (1- \gamma_{5} )  d^\beta ,  \nn \\ 
O_2 &=& \bar s^\alpha (1- \gamma_{5} ) d^\alpha \ 
 \bar s^\beta  (1- \gamma_{5} )  d^\beta ,  \nn \\ 
O_3&=& \bar s^\alpha  (1- \gamma_{5} )  d^\beta  \ 
 \bar s^\beta   (1- \gamma_{5} ) d^\alpha ,  \label{eq:ods2} \\ 
O_4 &=& \bar s^\alpha  (1- \gamma_{5} ) d^\alpha \  
\bar s^\beta  (1 + \gamma_{5} )  d^\beta ,  \nn \\ 
O_5&=& \bar s^\alpha  (1- \gamma_{5} )  d^\beta \ 
 \bar s^\beta (1 +  \gamma_{5} ) d^\alpha , \nonumber
 \eea 
where $\alpha$ and $\beta$ are color indices and $O^{\Delta 
S=2}\equiv O_{1}$.
For $\langle \bar K^0| O_i |K^0 \rangle$,  only the parity-even 
parts of the operators of eqs.~(\ref{eq:ods2})  contribute. 
$\langle \bar K ^0| O_1 |K^0 \rangle$ is expected
to vanish in the chiral limit, whereas the matrix elements 
$\langle\bar K^0| O_i |K^0 \rangle$ ($i=2,3,4,5$)
remain  finite. For the latter, close to the chiral limit,
we expect a mild dependence on the quark masses, as 
given in eq.~(\ref{eq:ecl}). 
\par Omitting terms which are of higher order in  
Chiral perturbation theory ($\chi$PT), the 
$B$-parameters are usually introduced using the expressions~\cite{NoiDELTAS=2}
\bea \langle  \bar K^{0} \vert    O_{1} (\mu) \vert K^{0} 
\rangle &=& \frac{8}{3} M_{K}^{2} f_{K}^{2} B_{1}(\mu) \, ,  \nn \\
\langle  \bar K^{0} \vert   O_{2} (\mu) \vert K^{0} \rangle  &=&
-\frac{5}{3} \left( \frac{ M_{K} }{ m_{s}(\mu) + m_d(\mu) }\right)^{2}
M_{K}^{2} f_{K}^{2} B_{2}(\mu)\, ,\nn   \\
\langle  \bar K^{0} \vert  O_{3} (\mu) \vert K^{0} \rangle  &=&
\frac{1}{3} \left( \frac{ M_{K} }{ m_{s}(\mu) + m_d(\mu) }\right)^{2}
M_{K}^{2} f_{K}^{2} B_{3}(\mu) \, ,\label{eq:VSAdef} \\
\langle  \bar K^{0} \vert   O_{4} (\mu) \vert K^{0} \rangle &=&
2  \left( \frac{ M_{K} }{ m_{s}(\mu) + m_d(\mu) }\right)^{2} 
M_{K}^{2} f_{K}^{2} B_{4}(\mu) \, ,\nn \\
\langle  \bar K^{0} \vert   O_{5} (\mu) \vert K^{0} \rangle  &=&
\frac{2}{3}\left( \frac{ M_{K} }{ m_{s}(\mu) + m_d(\mu) }\right)^{2}
 M_{K}^{2} f_{K}^{2} B_{5}(\mu) \nn \,  ,
 \eea
where $ O_{i}(\mu)$  and $m_{s}(\mu) + m_d(\mu)$ denote  operators 
and quark masses renormalized at the scale $\mu$. For the four-fermion 
operators the most common renormalization schemes are HV and 
NDR, although DRED is also occasionally used. 
\par In~(\ref{eq:VSAdef}), the matrix element of the operator $O_1$ is parameterized in terms of 
well-known experimental  quantities and $B_1(\mu)$ ($B_K(\mu)\equiv B_1(\mu)$)
has been computed with great precision on the 
lattice~\cite{shape1}--\cite{Sharpe96}.
The expression of the matrix elements $\langle \bar K^0| O_i |K^0 \rangle$ 
($i=2,3,4,5$) depends, instead, quadratically on the quark masses. Therefore, 
the ``conventional'' parameterization  introduces a redundant source
of  systematic error which can be avoided by parameterizing 
the matrix elements in terms of measured experimental 
quantities~\footnote{ Most recently,  $B$-parameters with the standard 
definitions~(\ref{eq:VSAdef})
have been computed   in  refs.~\cite{gupta_bp,NoiDELTAS=2}.}. 
\par  To overcome this problem, we  propose the following new parameterization of the $\Delta S =2$ operators 
\bea \langle  \bar K^{0} \vert    O_{1} (\mu) \vert K^{0} 
\rangle &=&
\frac{8}{3} M_{K}^{2} f_{K}^{2} B_1(\mu) ,  \nn \\
\langle  \bar K^{0} \vert    O_{i} (\mu) \vert K^{0} \rangle  &=&
M_{K^*}^{2} f_{K}^{2} \tilde{B}_i(\mu) \qquad (i=2,3,4,5)
\label{eq:FURBAdef}\; .
\eea
The advantage of~(\ref{eq:FURBAdef}) is that  it is still possible to 
work with dimensionless quantities, the $\tilde B_{i}(\mu)$s,  without 
any reference to the quark masses. In practice, one computes on the lattice
the ratios
\be \tilde B_{i} = \left( \frac{\bar K^{0} \vert   O_{i} (\mu) \vert K^{0} 
\rangle }{M_{K^*}^{2} f_{K}^{2}}\right)_{latt} \, ,\ee
with $i=2,3,4,5$, and derive the physical amplitudes, in GeV$^{4}$, using
\be\label{eq:phys}  
\langle \bar K^{0} \vert    O_{i} (\mu) \vert K^{0} \rangle=
\tilde B_{i}\left( M_{K^*}^{2} f_{K}^{2}\right)_{exp} \, .\ee
In sec.~{\ref{sec:numeri}} we use the new  parameterization to extract the matrix
elements of the operators $ O_i$ ($i=2,3,4,5$) from 
three-point correlation functions.
\par Another  possible  definition  is given  by the ratios
\be R_{i}= \left( \frac{\langle \bar K^{0} \vert   O_{i} (\mu) \vert K^{0} 
\rangle }{\langle \bar K^{0} \vert   O_{1} (\mu) \vert K^{0} 
\rangle} \frac{ M_{K}^{2} }{ M_{K^*}^{2} }\right)_{latt} \, 
,\label{eq:sdef}\ee
from which the physical matrix elements read
\be  \langle\bar K^{0} \vert    O_{i} (\mu) \vert K^{0} \rangle=
 \frac{8}{3} R_{i} B_{1}(\mu) \left( M_{K^*}^{2} f_{K}^{2}\right)_{exp} \, .\ee
In equation~(\ref{eq:sdef}) the factor $M_{K}^{2}/M_{K^{*}}^{2}$ has 
been introduced in order to have a smooth behavior of the ratios 
$R_{i}$ as a function of the quark masses.  \par 
The $R_{i}$ may be useful to directly  estimate the relative contribution 
to $ K^{0} $--$\bar K^{0}$  mixing coming from physics beyond the 
Standard Model~\cite{susy}--\cite{strumioski}:
\be \langle  \bar K^{0} \vert {\cal H}^{\Delta S=2} \vert K^{0} \rangle
\propto C_{1}(\mu) \langle \bar K^{0} \vert  O_{1} (\mu) \vert K^{0} 
\rangle \left(  1 + 
\frac{M^2_{K^*}}{M^2_K}\sum_{i=2,5} \frac{ C_{i}(\mu)}{ C_{1}(\mu)} R_{i} 
\right) \, \ee
where $ C_{1}(\mu)$ and $ C_{i}(\mu)$ are the Wilson coefficients of 
the corresponding operators. In the Standard Model, the coefficient $C_{1}$ is the only one 
different from zero. With our data, the error on 
$\langle \bar K^{0} \vert   O_{1} (\mu) \vert K^{0} 
\rangle$  is rather large. For this reason, we only present numerical results 
obtained  with the parameterization~(\ref{eq:FURBAdef}).
\item \underline{$\Delta I=3/2$ operators:}\\
The study of $\Delta I=3/2$
$K\rightarrow \pi \pi$  amplitudes requires the computation of the matrix elements 
$\langle\pi\pi| O^{3/2}_{i}|K\rangle$ of the  following left-left and left-right operators~\cite{epseM,epseR} 
\ba
O^{3/2}_{7} & = & 
{\bar s^\alpha} \gamma_{\mu}(1- \gamma_{5} ) d^\alpha\, 
{\bar u^\beta}  \gamma_{\mu}(1+ \gamma_{5} ) u^\beta + \left( d 
\leftrightarrow u \right) -  \left( \bar u  \to \bar  d,  u \to d  \right)  
 \nn \\
O^{3/2}_{8} & = & 
{\bar s^\alpha} \gamma_{\mu}(1- \gamma_{5} ) d^\beta\, 
{\bar u^\beta}  \gamma_{\mu}(1+ \gamma_{5} ) u^\alpha + \left( d 
\leftrightarrow u \right) -  \left( \bar u  \to \bar  d,  u \to d  \right) 
\label{eq:olri32}\\
O^{3/2}_{9}  & = &
{\bar s^\alpha} \gamma_{\mu}(1- \gamma_{5} ) d^\alpha\, 
{\bar u^\beta}  \gamma_{\mu}(1- \gamma_{5} ) u^\beta + \left( d 
\leftrightarrow u \right) -  \left( \bar u  \to \bar  d,  u \to d  \right)   
\, . \nonumber
\ea
These  matrix  elements are important  for 
the calculation of $\epsilon'/\epsilon$ in the 
Standard Model. 
In the chiral limit, the $\langle\pi\pi| O^{3/2}_{i}|K\rangle$  matrix elements can be  obtained,  using  soft pion theorems,  from  
$\langle\pi^{+}| O^{3/2}_{i}|K^{+}\rangle$ (to which only the parity-even parts of 
the operators contribute)
\be \langle\pi\pi| O_i|K\rangle_{I=2} =-\frac{1}{\sqrt{2}f_{\pi}} 
\langle\pi^{+}| O^{3/2}_{i}|K^{+}\rangle \, .\label{eq:soft}\ee
The latter can be computed  on the lattice 
using only the three-point correlation functions. For degenerate quark 
masses, $m_{s}=m_{d}=m$, and in 
the chiral limit,  we find
\ba\label{eq:softpion}
\chilim \langle\pi^{+}| O^{3/2}_{7}|K^{+}\rangle & = & - \chilim 
\langle \bar K^0| O_5 |K^0 \rangle = 
 - M_{\rho}^{2} f_{\pi}^{2} \chilim \tilde{B}_5(\mu)\nonumber\\
\chilim \langle\pi^{+}| O^{3/2}_{8}|K^{+}\rangle & = & - \chilim
\langle \bar K^0| O_4 |K^0 \rangle= 
 - M_{\rho}^{2} f_{\pi}^{2} \chilim \tilde{B}_4(\mu)\nonumber\\
\chilim \langle\pi^{+}| O^{3/2}_{9}|K^{+}\rangle & = &
\frac{4}{3} M_{\pi}^{2} f_{\pi}^{2} \chilim B_1(\mu)  \nonumber
\ea
\end{itemize}
where all the matrix elements are parameterized through well known 
experimental quantities. Thus we can predict the  
$\langle\pi^{+}| O^{3/2}_{i}|K^{+}\rangle$ matrix elements from the 
chiral limit of suitable $\tilde B$-parameters and from the chiral 
limit of $B_{1}$.
\vspace{-1.0cm}
\section{Renormalization Group Invariant Operators}
\label{sec:RGI}
\vspace{-0.5cm}
In this section, we give the main 
formulae which are necessary to define the Renormalization 
Group Invariant Wilson coefficients and operators for the most 
general $\Delta S =2$ effective Hamiltonian. The procedure is 
generalizable to any effective weak Hamiltonian.
\par Physical amplitudes can be written as 
\be 
\langle F \vert {\cal H}_{eff}\vert I \rangle = 
\langle F \vert \vec{ O}(\mu) \vert I \rangle \cdot \vec{C}(\mu) 
\, , \label{wope} 
\ee
where $\vec{ O}(\mu) \equiv
(  O_1(\mu),   O_2(\mu),   \dots,  O_N(\mu))$  is the operator basis, 
e.g. the basis  defined 
in~(\ref{eq:ods2}),  and  $\vec{C}(\mu)$  the corresponding Wilson coefficients
(see for example~\cite{ds2susy,strumioski})  represented as a column vector. 
$\vec C(\mu)$ is  expressed in terms of its counter-part, computed at a large scale $M$, 
through the renormalization-group evolution matrix  $\W[\mu,M]$
\be 
\vec C(\mu) = \W[\mu,M] \vec C(M)\, . \label{evo} 
\ee
\par The initial conditions for the evolution equations,
$\vec C(M)$, are obtained by perturbative matching 
of the full theory, which includes
propagating heavy-vector bosons ($W$ and $Z^0$), 
the top quark, SUSY particles, etc., to the effective theory where the $W$,
$Z^0$, the top quark and all the heavy particles have been integrated
out. In general,
$\vec C(M)$ depends on the scheme used to define the renormalized
operators. It is possible to show that $\W[\mu,M]$ can be 
written in the form 
\bea
\W[\mu,M]  = \hat M[\mu] \U[\mu, M] \hat M^{-1}[M] \, , 
 \label{monster} \eea
where $\U$ is the leading-order evolution matrix
\be
\U[\mu,M]=  \left[\frac{\as (M)}{\as (\mu)}\right]^{
      \gammaz_O / 2\beta_{ 0}} \, ,
\label{u0} \ee
and the NLO matrix is given by
\be \hat M[\mu] =
 \hat 1 +\frac{\as (\mu)}{4\pi}\J[\lambda(\mu)] \, .
\label{mo2} \ee
where $\gammaz_O$ is the leading order anomalous dimension matrix and 
$\J[\lambda(\mu)]$ is defined in~\cite{scimemi} and can be obtained 
by solving the Renormalization Group Equations (RGE) at the next-to-leading order.

The Wilson coefficients $\vec{C}(\mu)$ and the  renormalized operators 
$\vec{ O}(\mu)$ 
are usually defined  in a given   scheme ( HV, NDR, 
RI), at a fixed renormalization scale $\mu$, and depend on the 
renormalization scheme and scale. This is a source of confusion in 
the literature. Quite often, for example, one finds comparisons of 
$B$-parameters 
computed in different schemes. Incidentally, we note that the 
NDR scheme used in the lattice calculation of ref.~\cite{gupta_bp} 
differs from the standard NDR scheme of refs.~\cite{epseM,epseR}; on the other 
hand, the HV scheme of refs.~\cite{epseM} is not the same as the HV 
scheme of refs.~\cite{epseR}. In some cases, the differences between 
different schemes may be numerically large, e.g. $B_{8}^{(3/2) \HV}\sim  
1.3 \; B_{8}^{(3/2) \NDR}$.
For these reasons, the standard procedure is not entirely satisfactory, 
especially when the (perturbative) coefficients and the (non-perturbative) matrix elements 
are computed using different techniques, regularization, schemes and renormalization scales.
To avoid all these problems, we propose a Renormalization
Group Invariant (RGI) definition of Wilson coefficients and composite operators 
which generalizes what is usually done for $B_K$, by introducing the
RGI $B$-parameter $\hat B_K$ 
and for the quark masses~\cite{martiRGI,luscherRGI}.
The procedure is straightforward: from eq.~(\ref{u0}), we  define 
\be\label{eq:CRGI}
\hat{w}^{-1}[\mu] \equiv  \hat M[\mu] \left[\as (\mu) \right]^{
      - \gammaz_O / 2\beta_{ 0}}\, , 
\ee
which, using  eqs.~(\ref{monster}) and~(\ref{eq:CRGI}), gives
\be
\W[\mu,M]  = \hat{w}^{-1}[\mu]\hat{w}[M]\; .  
\ee
The effective Hamiltonian~(\ref{wope}) can then be written as 
\ba\label{eq:HRGIbella}
{\cal H}_{eff} 
 & = & \vec{ O}(\mu) \cdot       \vec C(\mu) 
    =     \vec{ O}(\mu)\hat W[\mu,M]  \vec C(M) \nonumber\\
 &   =    & \vec{ O}(\mu) \hat{w}^{-1}[\mu] \cdot \hat{w}[M] 
                                        \vec C(M)
    =     \vec{ O}^{RGI} \cdot \vec C^{RGI}\; ,  
\ea 
with
\ba\label{eq:monster2}
\vec C^{RGI}     & =  &\hat{w}[M] \vec C(M)\nonumber\\
\vec{ O}^{RGI}     & =   & \vec{ O}(\mu) \cdot \hat{w}^{-1}[\mu]\; .
\ea
$\vec C^{RGI}$ and  $\vec{ O}^{RGI}$ are scheme and scale independent at  
the order at which the Wilson coefficients have been computed
(NLO in most of the cases, NNLO for  quark masses). 
\par The $\tilde{B}$-parameters defined in eqs.~(\ref{eq:FURBAdef}) satisfy 
the same   renormalization group equations as the corresponding 
operators. The RGI $\tilde{B}$-parameters are then obtained 
from the relation
\be\label{eq:Brgi}
 {\tilde B}_i^{RGI} = \sum_{j} \tilde{B}_j(\mu)   w(\mu)_{ji}^{-1}\, .  \ee   
\vspace{-1.0cm}
\section{Numerical results}
\label{sec:numeri}
\vspace{-0.5cm}
In this section we present numerical results for 
the matrix elements of the basis~(\ref{eq:ods2}) and 
for the electro-penguin operators~(\ref{eq:olri32}),
obtained using  our new parameterization. 
\par 
All details concerning  the non-perturbative renormalization of  
lattice operators and  the extraction of matrix elements from 
correlation functions have been presented  elsewhere and 
are not repeated here.  The reader can find them, for example, 
in refs.~\cite{DELTAS=2,NoiDELTAS=2}, (see also ref.~\cite {4ferm_teo} for 
a complete discussion of the non-perturbative renormalization 
techniques  for $\Delta F=2$  operators and for references). In particular 
in~\cite{NoiDELTAS=2},  the same set of lattice data was analyzed 
and is compared here with the results  of  the our new study. 
\par     
Let us start by giving a simple argument which  
shows  the correlation existing between
the  values of the ``conventional'' $B$-parameters and the quark masses.
 \begin{table}[hptb]
\centering
\caption{\it{Matrix elements at the renormalization scale 
$\mu = a^{-1} \simeq 2$~GeV, corresponding to 
$\mu^{2} a^{2}=0.96$ and $\mu^{2} a^{2}=0.62$  at  
$\beta=6.0$ and $6.2$ respectively. All results are in 
the RI (MOM) scheme. In the first two columns
the  results of  the present  study obtained with the new 
parameterization are given. In the last two columns we 
show the results obtained 
with the ``conventional'' method in ref.~\cite{NoiDELTAS=2} on the 
same set of data (for $\langle O_1 \rangle$ the two parameterizations 
coincide).
$\langle  O^{3/2}_{7,8}\rangle$ stand for 
$\langle \pi^+| O^{3/2}_{7,8}|K^+ \rangle$
with $O^{3/2}_{7,8}$ given in eq.~(\ref{eq:olri32}).}}
\label{tab:summary}
\vspace{0.5cm}
\begin{tabular}{||c|c|cc||c|cc||}\hline\hline
      & \multicolumn{2}{c}{New}& &\multicolumn{2}{c}{Old}& \\
\hline
$\langle   O_i \rangle$ &$\beta=6.0$&$\beta=6.2$& 
& $\beta=6.0$  &$\beta=6.2$& \\
$\mu\simeq 2\mbox{GeV}$&this work& this work & &\cite{NoiDELTAS=2}
&\cite{NoiDELTAS=2}& \\
\hline \hline
$\langle   O_1 \rangle$& 0.012(2) & 0.011(3) & & 0.012(2) & 0.011(3) & \\ 
$B_1$   &                   0.70(15) & 0.68(21) & & 0.70(15) & 0.68(21) & \\
\hline
$\langle   O_2 \rangle$&-0.079(10)&-0.074(8) & &-0.073(15)&-0.073(15)& \\ 
$B_2$                     & 0.72(9)  &0.67(7)   & & 0.66(3)  & 0.66(4)  & \\
\hline
$\langle   O_3 \rangle$&0.027(2)  & 0.021(3) & &0.025(5)  &0.022(5)  & \\ 
$B_3$                     &1.21(10)  & 0.95(15) & & 1.12(7)  & 0.98(12) & \\
\hline
$\langle  O_4 \rangle$& 0.151(7) & 0.133(12)& &0.139(28) &0.133(28) & \\ 
$B_4$                     & 1.15(5)  & 1.00(9)  & & 1.05(3)  & 1.01(6)  & \\
\hline
$\langle   O_5 \rangle$& 0.039(3) & 0.029(5) & &0.035(7)  &0.029(7)  & \\ 
$B_5$                     & 0.88(6)  & 0.66(11) & & 0.79(6)  & 0.67(10) & \\
\hline
$\langle   O^{3/2}_7 
\rangle$                  &-0.019(2) &-0.011(3) & &-0.020(5) &-0.014(5) & \\ 
$B^{3/2}_7$               & 0.65(5)  & 0.38(11) & & 0.68(7)  & 0.46(13) & \\
\hline
$\langle   O^{3/2}_8 
\rangle$                  &-0.082(4) &-0.068(8) & &-0.092(19)&-0.087(19)& \\ 
$B^{3/2}_8$               & 0.92(5) &0.77(9) & & 1.04(4) & 0.98(8) & \\
\hline \hline
\end{tabular}
\end{table}
In order to extract the operator matrix elements  the following two- and 
three-point correlation functions are used
\bea
G_P(t_x,\vec p) = \sum_{\vec x} 
\langle P(x) P^{\dagger} (0) \rangle e^{-\vec p \cdot \vec x} &,&
\,\,\,\,\,\,\,\,\,\,\,
G_A(t_x,\vec p) = \sum_{\vec x} \langle A_0(x) P^{\dagger} (0) 
\rangle e^{-\vec p 
\cdot \vec x}, 
\nonumber\\
G_{ O}(t_x,t_y;\vec p, \vec q) &=& \sum_{\vec x,
\vec y} \langle P^{\dagger}(y) O(0) P^{\dagger}(x)\rangle
e^{-\vec p \cdot \vec y} e^{\vec q \cdot \vec x} , 
\label{eq:corrs} 
\eea
where $x \equiv (\vec x, t_x), y \equiv (\vec y , t_y)$ and
$P$, $A_{0}$  and $ O$ stand for the  renormalized pseudoscalar 
density, the fourth component of the axial current and four-fermion operator 
respectively~\footnote{ In this study, 
all correlation functions  are evaluated with degenerate quark
masses.}.
\begin{figure}[tbh]
\caption{\it{$\langle  \bar K^{0} \vert  O_{4} \vert K^{0} \rangle$ 
in GeV$^4$ computed with the new parameterization as a 
function of the sum of the quark masses $m_s + m_d$ 
renormalized in the $\msbar$ scheme at $\mu=\simeq 2 $~GeV.
The quark masses span a range of values compatible with 
theoretical estimates.}}
\label{fig:fig.1}
\ifig{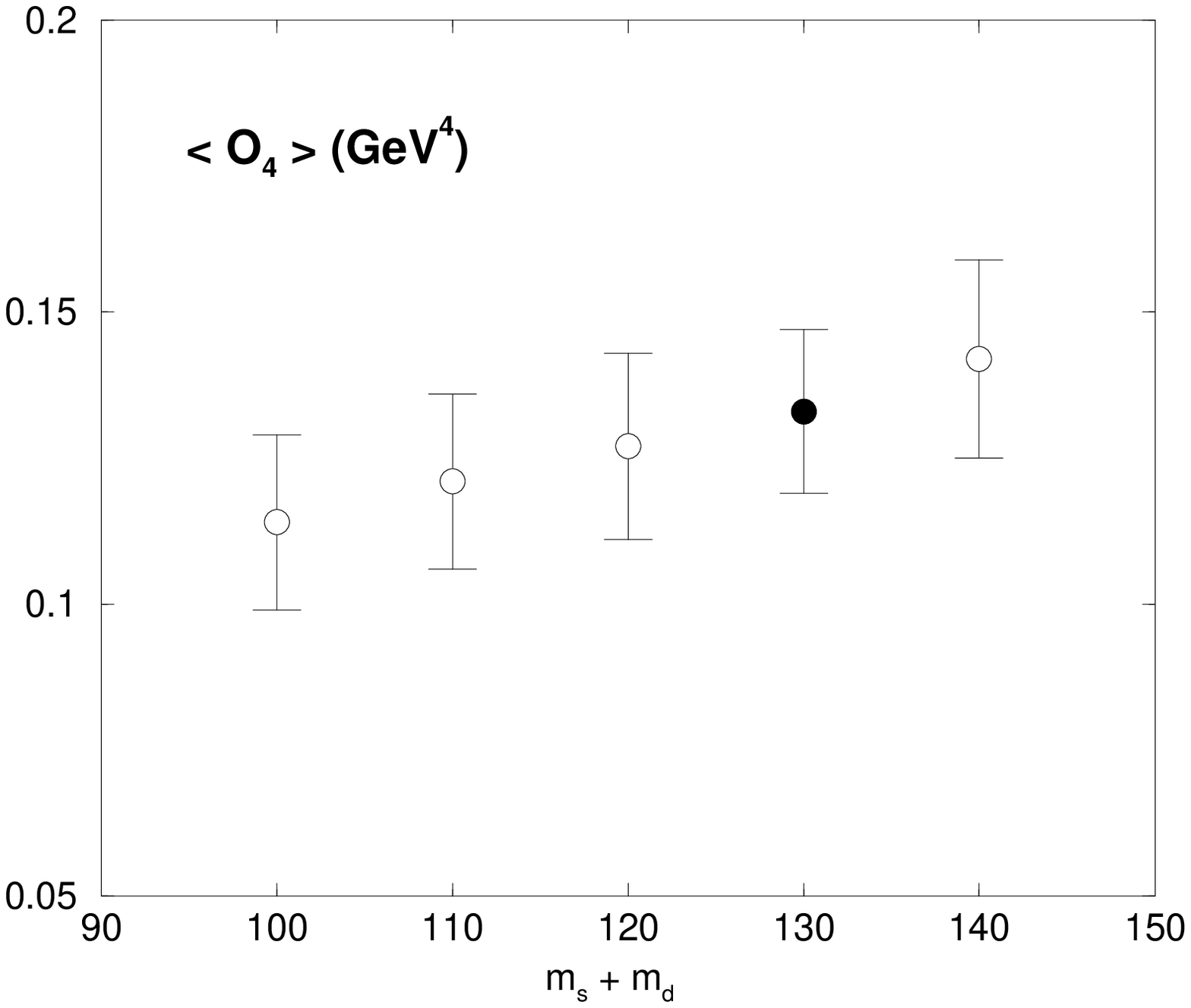}
\end{figure} 
By forming suitable ratios of the above correlations
and looking at their asymptotic behavior at large time separations, 
one can directly obtain  the $B$-parameters defined in eqs.~(\ref{eq:VSAdef}). 
At the leading order in the chiral expansion, or using the 
definitions of ref.~\cite {NoiDELTAS=2}, one finds (for $i=2,3,4,5$) 
\bea
 R_i & =& 
                \frac{G_{ O_i }(t_{x}, t_{y}, \vec p,\vec q)}{ G_P(t_{x}, 
                \vec p) G_P(t_{y}, \vec q)} 
             \to \frac{\langle \bar K^0(\vec q)               
           \vert {O}_i \vert K^0(\vec p) \rangle}{
   \vert \langle 0 \vert P \vert K^0 \rangle \vert ^2 }=  
 \mbox{const.} \times B_{i}(\mu) \, .
\label{eq:rapporti} 
\eea
On the other hand, quark masses are extracted using vector or 
axial-vector Ward identities, e.g.
\bea
 m_{s}(\mu)+m_{d}(\mu) & =&  \frac{\langle 0             
           \vert \partial_{\nu}  A_{\nu}  \vert K^0\rangle}{
   \langle 0 \vert  P \vert K^0 \rangle  }\, ,
\label{eq:axwi} 
\eea
where the quark masses $m_{s}(\mu)+m_{d}(\mu)$ and the renormalized pseudoscalar 
density $ P$  are given, by definition, in the same renormalization 
scheme.  This happens because matrix elements of the ``good'' axial 
currents, and consequently the product $(m_{s}(\mu)+m_{d}(\mu))  P$ are 
regularization, renormalization scheme and scale independent.
In practice, for a given renormalization scheme and scale adopted to 
renormalize $ P$, the matrix elements 
$\langle 0  \vert \partial_{\nu}  A_{\nu}  \vert K^0\rangle$
and $\langle 0 \vert  P \vert K^0 \rangle$  are computed numerically 
and from their ratio one obtains $m_{s}(\mu)+m_{d}(\mu)$. 
\par From a comparison of eqs.~(\ref{eq:rapporti}) and~(\ref{eq:axwi}), it 
is evident that, for  large values of 
 $\langle 0 \vert  P \vert K^0 \rangle$,  
small values of the quark masses  and  of the $B$-parameters will be 
simultaneously obtained.  The ratio
$B_{i}/((m_{s}(\mu)+m_{d}(\mu))^{2}$ , corresponding to the 
``physical'' matrix elements, including the mass factors appearing 
in  eqs.~(\ref{eq:VSAdef}),  will  however 
be much less dependent on the quark masses since $B_{i}$ 
and $m_{s}(\mu)+m_{d}(\mu)$
are strongly correlated, i.e. 
$\langle 0 \vert  P \vert K^0 \rangle\propto1/m_q$. 
It would be necessary 
that calculations of  ``conventional'' $B$-parameters  provide 
at the same time the value of the quark masses obtained in the same 
calculation. This is equivalent to give, as suggested 
in eq.~(\ref {eq:FURBAdef}), the matrix elements in physical units.
\par
In this work we have used the lattice correlation functions computed 
in~\cite{NoiDELTAS=2}. They have  been obtained from numerical simulations 
at $\beta = 6.0$ 
(460 configurations, Run A) and   $6.2$  (200 configurations, Run B) with the tree-level Clover
action, for several values of the quark masses and for different 
meson momenta.   We have used    the ``lattice dispersion 
relation'' $\sinh^{2} \left(\frac{E(\vec p)}{2}\right) = \sinh^{2}
\left(\frac{m}{2}\right)  +\sum_{i=1,3} \sin^{2} 
\left(\frac{p_{i}}{2} \right)$. The physical volume is 
approximatively the same on the two lattices. Statistical errors have been estimated with the 
jackknife method.  
\par The main numerical results that we have obtained are listed in 
table~\ref{tab:summary}. They have been computed in the RI(MOM) scheme 
at $\mu=2$~GeV as follows:
\begin{itemize}
\item In the first two columns of the table the results, 
obtained with 
the new parameterization, are given. The $\tilde B$-parameters are 
obtained by extrapolating/interpolating the ratios
\be\label{eq:ultima}
R'_i = \frac{M^2_P}{M^2_V}
                \frac{G_{ O_i }(t_{x}, t_{y}, \vec p,\vec q)}{ G_A(t_{x}, 
                \vec p) G_A(t_{y}, \vec q)} 
             \to \frac{\langle \bar K^0(\vec q)               
           \vert {O}_i \vert K^0(\vec p) \rangle}{
   M_V^2 f^2_P} \; .
\ee
to the physical point using the lattice-plane method~\cite{lpmethod}. 
$M_P$ and $M_V$ are the masses of the pseudoscalar and vector mesons 
computed at the same quark masses as the correlation functions. 
Then the matrix elements $\langle  O_i \rangle$ in GeV$^4$ are obtained 
from eq.~(\ref{eq:phys}). The standard $B_i$'s given in 
these columns are obtained from eqs.~(\ref{eq:VSAdef}) 
by using the matrix elements in GeV$^4$ in the same column and 
a ``conventional'' quark mass 
$(m_s+m_d)^{\overline{MS}}(2\; GeV) = 130$~MeV. This is 
the value of the strange quark mass 
that we have obtained on the same sets of 
data~\cite{Bello}.
\item   In the last two columns we show the results obtained 
with the ``conventional'' method in ref.~\cite{NoiDELTAS=2} on the 
same set of data. In this case we first obtain the  
$B$-parameters,  
defined as in eq.~(\ref{eq:VSAdef}), using the procedure 
described in~\cite{NoiDELTAS=2}. The matrix elements in GeV$^4$
reported in these columns are derived from the "conventional" B-parameters 
in eqs.~(\ref{eq:VSAdef}) with 
the same quark as before of 
$(m_s+m_d)^{\overline{MS}}(2\; GeV) = (130 \pm 15)$~MeV. In this case we have 
to include the uncertainty coming from the error in the determination of 
the quark mass. The errors 
on the $B$-parameters and the quark 
masses are considered as independent to mimic what is 
usually done in phenomenological analyses. 
\end{itemize} 
The results in table~\ref{tab:summary} show the convenience of the 
approach proposed in this paper. 
Most phenomenological analyses, which 
use the standard procedure, adopt values of $B$-parameters and 
quark masses taken from different determinations. For example
the $B$-parameter is taken from a calculation at a fixed value of the 
lattice spacing in the quenched approximation whereas
the quark mass is taken from some compilation of lattice 
results extrapolated to the continuum and including some 
evaluation of the quenching errors. In other cases,
the value of the $B$-parameter is a sort of average
biased by results from the large $N$ approach and 
lattice calculations and the mass is taken from an average 
from lattice and QCD sum rules. In this way, one misses the correlation
between the value of the $B$-parameter and the mass. 
As a result, the uncertainty on the physical matrix 
element is much larger. In figure~\ref{fig:fig.1} we present   
the values of the matrix element 
$\langle  \bar K^{0} \vert  O_{4} \vert K^{0} \rangle$ 
obtained with the new parameterization 
for different  choices of $m_s+m_d$,
in a range of values compatible with theoretical estimates.
To be more specific we have computed 
$\langle  \bar K^{0} \vert  O_{4} \vert K^{0} \rangle$
for different values of $m_s + m_d$, 
where $m_s$ and $m_d$ are the masses
of the quarks in the meson. This means that different 
choices of $m_s + m_d$ also correspond to different values 
of $M_{K^0}$ (and of $f_K$). \\
Although  we have data at two different values of the lattice spacing, 
the 
statistical errors, and the uncertainties in the extraction of the 
matrix elements,  are too large to enable any extrapolation to the continuum limit 
$a \to 0$ : within the precision of our results  we cannot detect the dependence of
$\tilde B$-parameters on $a$. For this reason,  we estimate the central 
values by  averaging the $B$-parameters obtained with the physical mass
$M_K^{exp}$ at the two values of 
$\beta$. Since the results at $\beta=6.0$ have  smaller statistical 
errors but suffer from larger discretization  effects, we do not 
weight the averages with the quoted statistical 
errors but simply take the sum of the two values divided by two. As 
far as the errors are concerned we take the largest of the two statistical 
errors. This is a rather conservative way of estimating the 
errors.  In order to compare the results of Run A and Run B, we have 
chosen the same 
physical renormalization scale  $\mu$. 
Using  estimates of the lattice spacing ($a^{-1} = 2.12(6)$ at $\beta=6.0$ and 
$a^{-1} = 2.7(1)$ at $\beta=6.2$) of ref.~\cite{Bello},
we have taken $\mu^{2} a^{2}=0.96$ and $\mu^{2} 
a^{2}=0.62$, corresponding to $\mu=2.08$ GeV and $\mu=2.12$ GeV, at 
$\beta=6.0$ and $6.2$ respectively.  We quote the results as obtained 
at $\mu=2$ GeV, since the running of the matrix elements between $\mu 
\sim 2.1$ and $2.0$ is totally negligible in comparison with the 
final errors.
\vspace{-0.8cm} 
\section{Physical Results}
\label{sec:fine}
\vspace{-0.5cm}
Our best estimates of the matrix elements
of $\Delta S =2 $ operators in the RI scheme at $\mu=2$~GeV are given 
in table~\ref{tab:summarybest}.
Note that these matrix elements  
are enhanced 
by a factor $\simeq 2 \div 12$ with respect to the SM one 
($\langle  O_1 \rangle$). For this reason, $K^0-\bar{K}^0$ mixing 
is a promising observable to detect signals of new physics at low 
energy~\cite{ds2susy,strumioski}.\\
From the matrix elements obtained non-perturbatively in the 
RI scheme, we 
derive the RGI $\tilde B$-parameters 
using continuum perturbation theory at NLO. These 
have been computed from eqs.~(\ref{eq:Brgi}) with 
$\alpha_s(2 \mbox{GeV})=0.31$, corresponding to $\alpha_s(M_Z)=0.118$, 
evaluated with the appropriate number of active flavors ( 
$n_f=4$ at 2 GeV).
The results are given in table~\ref{tab:summarybest}.
This choice mimics what is usually done in 
phenomenological analyses which use lattice QCD estimates. It 
corresponds to the assumption that the results 
in the RI-scheme are the "physical" matrix elements, up to
some undetermined unquenching errors.
 \begin{table}[htb]
\centering
\caption{\it{Matrix elements in GeV$^4$ at $\mu=2$~GeV in the RI scheme and 
their RGI values with $\alpha_s^{n_f=4}$.}}
\label{tab:summarybest}
\begin{tabular}{||c|c|c||}\hline\hline
$\langle   O_i\rangle$ & RI & RGI \\
\hline \hline
$\langle   O_1 \rangle$ & 0.012(3)   & 0.017(4) \\ 
\hline
$\langle   O_2 \rangle$ & -0.077(10) & -0.050(7)\\ 
\hline
$\langle   O_3 \rangle$ & 0.024(3)   & 0.001(7)\\ 
\hline
$\langle   O_4 \rangle$ & 0.142(12)  & 0.068(6)\\ 
\hline
$\langle   O_5 \rangle$ & 0.034(5)   & 0.038(5)\\ 
\hline \hline
\end{tabular}
\end{table}
\par Our best estimate of the matrix elements of $\Delta I =3/2$ operators
in the RI scheme at $\mu=2$~GeV are 
\ba\label{eq:3/2}
\langle \pi^+| O_{7}^{3/2}|K^+\rangle & = & -(0.015 \pm 0.004) 
\mbox{GeV}^4\; ,\nonumber \\
\langle \pi^+| O_{8}^{3/2}|K^+\rangle & = & -(0.075 \pm 0.008)\mbox{GeV}^4\; . 
\ea
Since the analyses of 
refs.~\cite{nonper}, \cite{ciuc3}-\cite{paschoslast} 
are done using Wilson coefficients
in NDR and HV, by using the  matching coefficients between the RI and these schemes~\cite{nonper}
\be\label{eq:copiata}
\left ( \hat O_i^{3/2} \right )^{\NDR,\HV} =
\left ( \delta_{ij} - \frac{\alpha_s(\mu)}{4\pi} \Delta r_{ij}^{\NDR,\HV} \right )
\left ( \hat O_j^{3/2} \right )^{\RI}
\ee
where
\[
\hspace{-0.9cm}
\Delta r_{ij}^{\NDR} = \left (
\begin{array}{cc} 
\frac{2}{3} + \frac{2}{3} \ln 2 & -2 -2 \ln 2 \\
2 - 2 \ln 2 & - \frac{34}{3} + \frac{2}{3} \ln 2 
\end{array}
\right ) \;\;
\Delta r_{ij}^{\HV} = \left (
\begin{array}{cc} 
-\frac{8}{3} + \frac{2}{3} \ln 2 & -8 -2 \ln 2 \\
-2 - 2 \ln 2 & - \frac{62}{3} + \frac{2}{3} \ln 2 
\end{array}
\right ). 
\]
and the results in (\ref{eq:3/2}), we have computed the matrix elements 
$\langle \pi^+ | O^{3/2}_i|K^+ \rangle$ given in 
table~\ref{tab:summaryNDRHV}.
 \begin{table}[t]
\centering
\caption{\it{Matrix elements in GeV$^4$ at $\mu=2$~GeV in the NDR and HV 
schemes and with $\alpha_s^{n_f=4}(2\mbox{GeV})=0.31$.}}
\label{tab:summaryNDRHV}
\begin{tabular}{||c|c|c||}\hline\hline
$\langle   O^{3/2}_i\rangle$ & NDR & HV \\
\hline \hline
$\langle \pi^+ | O^{3/2}_7|K^+ \rangle$      & -0.021(4)  & -0.033(5) \\ 
\hline
$\langle \pi \pi| O_7|K \rangle_{I=2}$ & 0.11(2)   & 0.18(3) \\ 
\hline\hline
$\langle \pi^+ | O^{3/2}_8|K^+ \rangle$      & -0.095(10) & -0.114(12) \\ 
\hline
$\langle \pi \pi| O_8|K \rangle_{I=2}$ & 0.51(5)    & 0.62(6) \\ 
\hline \hline
\end{tabular}
\end{table}
To obtain $\langle \pi \pi| O_i|K \rangle_{I=2}$ we have used 
the chiral relation of eq.~(\ref{eq:soft}). This entails further 
uncertainty in the numerical evaluation of the physical matrix elements: 
within our accuracy, we may use in eq.~(\ref{eq:soft}) 
$f_K$ instead of $f_\pi$.
Moreover, to obtain the matrix elements in the $\overline{MS}$ 
scheme from those obtained 
non-perturbatively in the RI-scheme, we have chosen 
the unquenched value $\alpha_s(2\mbox{GeV})=0.31$ but, within 
the quenched approximation, we could have chosen the quenched 
value of $\alpha_s$ as well. We estimate
that, due to these effects, the final error is about twice that quoted 
in table~\ref{tab:summaryNDRHV}.\\
It is interesting to compare our result for 
$\langle \pi^+ | O^{3/2}_7|K^+ \rangle$ in the NDR scheme
with the recent estimate from large 
$N$ expansion \cite{perito}. 
The two determinations differ by more than two $\sigma$
with respect to the error quoted in table~\ref{tab:summaryNDRHV}. 
\vspace{-0.8cm}
\section{Conclusions}
\label{sec:conclusions}
\vspace{-0.5cm}
In this work we have introduced a new parameterization of four fermion 
operator matrix elements which does not involve quark masses
and thus allows a reduction of systematic uncertainties.
As a result the apparent quadratic dependence of 
$\epsilon'/\epsilon$ on the strange quark mass is removed.
We have also defined Renormalization Group 
Invariant matrix elements to simplify the matching between the lattice 
and continuum renormalization schemes. We have used these definitions 
to compute matrix elements of $\Delta I =3/2$ and SUSY $\Delta S =2$
four fermion operators on the lattice in the quenched approximation. 
The simulations have been performed at two different values of the lattice 
spacing and the renormalization constants of the operators are 
calculated non perturbatively.
\vspace{-0.8cm}
\section*{Acknowledgments}
\vspace{-0.5cm}
We thank M.~Ciuchini, L.~Conti, E.~Franco, V.~Lubicz 
and A.~Vladikas for interesting discussions.
A. D. acknowledges the I.N.F.N. for financial support.
V. G. has been supported by CICYT under the Grant AEN-96-1718,
by DGESIC under the Grant PB97-1261 and by the Generalitat Valenciana
under the Grant GV98-01-80. L. G. has been supported in part 
under DOE grant DE-FG02-91ER40676. G. M. acknowledges the M.U.R.S.T.
and the INFN for partial support.


\vspace{-0.8cm}
\begin{thebibliography}{999}
\vspace{-0.5cm}
%
\def\edlat{Lattice 97, 15th Int. Symp. on Lattice Field Theory,
Edinburgh, Scotland, 1997}
\def\stlolat{Lattice 96, 14th Int. Symp. on Lattice Field Theory, St
Louis, USA, 1996}
\def\ozlat{Lattice 95, 13th Int. Symp. on Lattice Field Theory,
Melbourne, Australia, 1995}
\def\biellat{Lattice 94, 12th Int. Symp. on Lattice Field Theory,
Bielefeld, Germany, 1994}
\def\txlat{Lattice 93, 11th Int. Symp. on Lattice Field Theory,
Dallas, Texas, 1993}
\def\nllat{Lattice 92, 10th Int. Symp. on Lattice Field Theory,
Amsterdam, Netherlands, 1992}
\def\warsaw{ICHEP96, 28th Int. Conf. on High Energy Physics, Warsaw,
Poland, 25--31 July 1996, edited by Z. Ajduk and A.K. Wroblewski,
World Scientific, Singapore (1997)}
%
\def\prd#1{Phys. Rev. D {\bf #1}}
\def\prl#1{Phys. Rev. Lett. {\bf #1}}
\def\plb#1{Phys. Lett. B {\bf #1}}
\def\npb#1{Nucl. Phys. B {\bf #1}}
\def\npbps#1{Nucl. Phys. B (Proc. Suppl.) {\bf #1}}
\def\npaps#1{Nucl. Phys. A (Proc. Suppl.) {\bf #1}}
\def\zpc#1{Z. Phys C {\bf #1}}
\def\nima#1{Nucl. Instrum. Meth. A {\bf #1}}
\def\cmp#1{Commun. Math. Phys. {\bf #1}}
\def\physrep#1{Phys. Rep. {\bf #1}}
%
\bibitem{CMP}
N. Cabibbo, G. Martinelli and R. Petronzio, Nucl. Phys. B244 (1984) 381.
\bibitem{BGGM}
R.C. Brower, M.B. Gavela, R. Gupta and G. Maturana,\\
 Phys. Rev. Lett. 53 (1984) 1318.
\bibitem{bernardargo} C.~Bernard, Argonne 1984, Proceedings of Gauge 
Theory on a Lattice,\\ 
p.85-101, UCLA-84-TEP-03.
\bibitem{boc}
M.~Bochicchio et al., Nucl. Phys. B262 (1985) 331.
\bibitem{shape1} S.~Sharpe et al., Nucl.~Phys. B286 (1987) 253.
\bibitem{ksgp} G.~Kilcup, S.~Sharpe, R.~Gupta and A.~Patel,
Phys.~Rev.~Lett. 64 (1990) 25.
\bibitem{jap}
N.~Ishizuka et al., Nucl. Phys. B (Proc. Suppl.) 34 (1994) 403;\\
Phys. Rev. Lett. 71 (1993) 24.
\bibitem{jap1}
S.~Aoki et al., Phys.~Rev.~Lett.80 (1998) 5271
\bibitem{Sharpe96}
S.R.~Sharpe, Nucl. Phys. B (Proc. Suppl.) 53 (1997) 181; R.~Gupta,
Invited talk given at 16th Autumn School and Workshop on Fermion Masses, 
Mixing and CP Violation (South European Schools on
Elementary Particle Physics) (CPMASS 97), Lisbon, Portugal, 6-15 Oct 1997, 
hep-ph/9801412.\\
L. Lellouch (CERN), 
Invited talk at 34th Rencontres de Moriond: 
Electroweak Interactions and Unified Theories, Les
Arcs, France, 13-20 Mar 1999, hep-ph/9906497. 
\bibitem{bs}
C.~Bernard et al., Nucl.~Phys.~B (Proc. Suppl.)~4~(1988)~483.
\bibitem{franco} E.~Franco et al., Nucl.~Phys. {\bf B317} (1989) 63.
\bibitem{GAVELA} M. B. Gavela et al., Nucl. Phys. B 306 (1988) 677;\\
 Nucl. Phys. B (Proc. Suppl.) 17 (1990) 769.
\bibitem{cbas} C.~Bernard and A.~Soni, Nucl.~Phys.~B~(Proc. 
Suppl.)~17~(1990)~495;\\  
Nucl. Phys.~B~(Proc. Suppl.)~42~(1995)~391.
\bibitem{DELTAS=2} G.~Martinelli et al., Nucl.~Phys.~B445~(1995)~81; 
A.~Donini et al.,  Phys.~Lett.~B360~(1996)~83;
M.~Crisafulli et al., Phys.~Lett.~B369~(1996)~325;
L. Conti et al.,  Phys.~Lett.~B421~(1998)~273.
\bibitem{gupta_bp}  T.~Bhattacharya, R.~Gupta and S.~Sharpe, 
Phys.~Rev. {\bf D55} (1997) 4036.
\bibitem{NoiDELTAS=2} C.~R.~Allton et al.,
Phys. Lett. B 453 (1999)30. 
\bibitem{soni}
G.~Kilcup,
Nucl. Phys. B (Proc. Suppl.) 20 (1991) 417;\\
S. Sharpe,
Nucl. Phys. B (Proc. Suppl.) 20 (1991) 429;\\
T.~Blum et al.,
hep-lat/9808025.
\bibitem{ds2susy} 
M.~Ciuchini et al.,
\newblock JHEP 9810:008,1998. 
\bibitem{epseM} 
A. Buras, M. Jamin, M. E. Lautenbacher, 
Nucl. Phys. B 408 (1993) 209. 
\bibitem{epseR}
M. Ciuchini, E. Franco, G. Martinelli, L. Reina,
Nucl. Phys. B 415 (1994) 403. 
\bibitem{oon} 
W.A.~Bardeen, A.J.~Buras and J.M.~G\'erard, Phys. Lett. B 180 
(1986) 133; Nucl. Phys. B 293 (1987) 787; Phys. Lett. B 192 (1987) 138.\\
E.A.~Paschos and Y.L.~Wu,  Mod. Phys. Lett. A6 (1991) 93.
\bibitem{qmasses} 
R.~D.~Kenway, 
Nucl. Phys. (Proc.Suppl.) 73 (1999)~16 and reference therein.\\ 
T. Bhattacharya and R. Gupta,
Nucl. Phys. (Proc.Suppl.) 63 (1998)95.\\
S. R. Sharpe, 
Talk given at 29th International Conference on High-Energy Physics (ICHEP 98), 
Vancouver, Canada 1998, HEP Vol.~1 p.~171.\\
V. Lubicz, 
Nucl. Phys. (Proc.Suppl.) 74 (1999) 291. 
\bibitem{NP} G.~Martinelli, C.~Pittori, C.T.~Sachrajda, M.~Testa and 
A.~Vladikas,\\ 
Nucl. Phys.~B445~(1995)~81.
\bibitem{nonper} M.~Ciuchini et al., Z. Phys. C68 (1995) 239.
\bibitem{susy}
F. Gabbiani, E. Gabrielli, A. Masiero and L. Silvestrini,\\ 
Nucl. Phys.~B~477~(1996)~321.
\bibitem{susybag} 
J. Bagger, K. Matchev and R. Zhang, 
\newblock Phys. Lett.~B 412~(1997)~77.
\bibitem{strumioski}
L.~Giusti, A.~Romanino and A.~Strumia , 
Nucl. Phys. B 550 (1999)3.
\bibitem{scimemi}  M. Ciuchini et al., Nucl. Phys.~B~523~(1998)~501. 
\bibitem{martiRGI}
A. Gonzalez Arroyo, F.J. Yndurain and G. Martinelli,\\
Phys. Lett. B 117~(1982)~437; Erratum-ibid.  B~122~(1983)~486. 
\bibitem{luscherRGI} S.~Capitani et al., Nucl. Phys.~B~544~(1999)~669.  
\bibitem{4ferm_teo}
A.~Donini, V.~Gimenez, G.~Martinelli, M.~Talevi and A.~Vladikas,\\
Eur. Phys. J. C 10 (1999)~121.
\bibitem{lpmethod}
C.R. Allton, V. Gimenez, L. Giusti, F. Rapuano,
Nucl. Phys. B 489 (1997)427. 
\bibitem{Bello}
V. Gimenez, L. Giusti, F. Rapuano, M. Talevi,
Nucl. Phys. B 540 (1999)472. 
\bibitem{ciuc3} M.~Ciuchini, Nucl. Phys.  (Proc. Suppl.) 59 (1997) 149.
\bibitem{buras2} A. Buras, M. Jamin and  M.E. Lautenbacher, 
Phys. Lett.  B 389 (1996) 749.
\bibitem{bert1} S.~Bertolini, J.O. Eeg and M.~Fabbrichesi,  
Nucl. Phys.    B 476  (1996) 225.
\bibitem{bert2} S.~Bertolini, J.O. Eeg, M.~Fabbrichesi and E.I.~Lashin, 
 Nucl. Phys.   B 514 (1998) 93. 
\bibitem{silvestrini} S.~Bosch {\em et al.}, hep-ph/9904408.
\bibitem{paschoslast} T.~Hambye, G.O.~K\"ohler, E.A.~Paschos and 
P.H.~Soldan, hep-ph/9906434.
\bibitem{perito}
M. Knecht, S. Peris and E. de Rafael 
Phys. Lett. B 457~(1999)~227.
\end{thebibliography}
\end{document}